\documentclass[a4paper,11pt]{article}
\usepackage[english]{babel}
\usepackage{pbox}
\usepackage[utf8]{inputenc}
\usepackage{csquotes}
\usepackage{amsmath}
\usepackage{graphicx}
\usepackage{cite}
\usepackage%[margin=2.5cm]
{geometry}
\usepackage[bf, font={small}]{caption}
\usepackage[plain]{fancyref}
\usepackage{authblk}

\usepackage{amstext} % for \text macro
\usepackage{array}   % for \newcolumntype macro
\newcolumntype{L}{>{$}l<{$}} % math-mode version of "l" column type
\usepackage{siunitx}

\title{\bf Microresonator Isolators and Circulators Based on the Intrinsic Nonreciprocity of the Kerr Effect}
\author[1,2]{Leonardo~Del~Bino$^\dagger$}
\author[1]{Jonathan~M.~Silver$^\dagger$}
\author[1,2]{Michael~T.~M.~Woodley}
\author[1]{Sarah~L.~Stebbings}
\author[1,3]{Xin~Xhao}
\author[1,*]{Pascal~Del'Haye}
\affil[1]{National Physical Laboratory (NPL), Teddington, TW11 0LW, United Kingdom}
\affil[2]{Heriot-Watt University, Edinburgh, EH14 4AS, Scotland}
\affil[3]{School of Electronic and Information Engineering, Beihang University, Beijing 100083, China}
\affil[*]{Corresponding Author: pascal.delhaye@npl.co.uk}
\date{}

\begin{document}
\maketitle

\begin{abstract}
Nonreciprocal light propagation is important in many applications, ranging from optical telecommunications to integrated photonics. A simple way to achieve optical nonreciprocity is to use the nonlinear interaction between counterpropagating light in a Kerr medium. Within a ring resonator, this leads to spontaneous symmetry breaking, with the result that light of a given frequency can circulate in one direction, but not in both directions simultaneously. In this work, we demonstrate that this effect can be used to realize optical isolators and circulators based on a single ultra-high-Q microresonator. We obtain isolation of $>$24 dB and develop a theoretical model for the power scaling of the attainable nonreciprocity.
\end{abstract}
\begin{figure}[h]
    \centering
    \includegraphics[width=\textwidth]{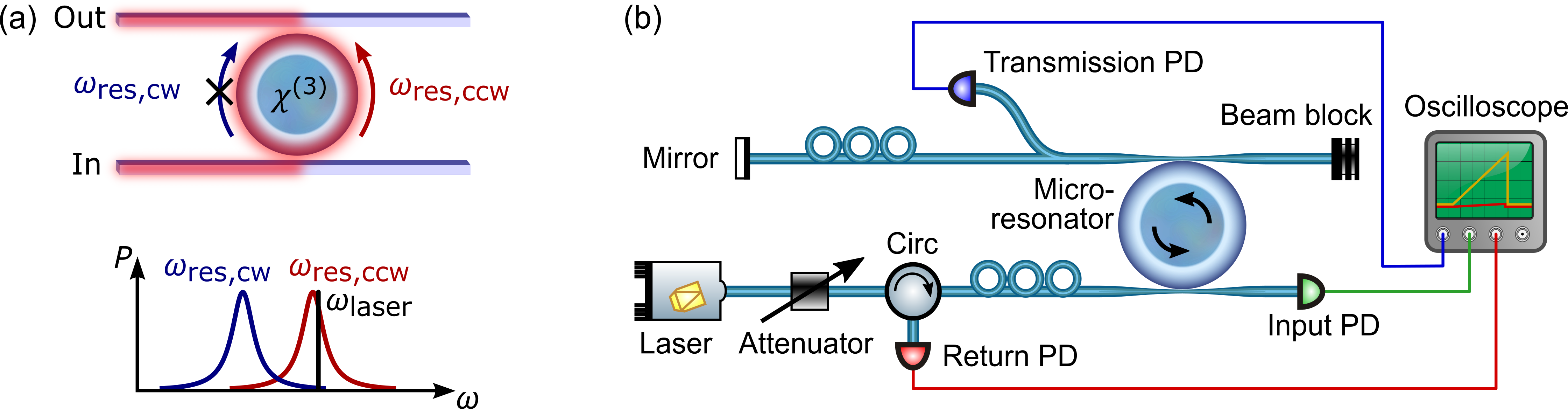}
    \caption{Kerr nonreciprocity in a microresonator and experimental setup for an isolator. (a) Schematic of the nonreciprocity.  Light is coupled in counterclockwise (ccw) direction through an ultra-high-Q microresonator in an add-drop configuration. The presence of the ccw light induces a Kerr shift that is twice as strong for clockwise circulating light, which leads to a resonance frequency splitting (shown in the lower panel). As a consequence, backwards propagating light (clockwise, cw) cannot pass through the resonator. (b) Experimental setup for the characterization of an isolator based on the Kerr nonreciprocity. Laser light is coupled through an ultra-high-Q silica microrod resonator using two tapered optical fibers. A fibre mirror simulates an optical circuit that reflects $\sim$100\% of the incident light and the polarisation is adjusted to maximise the power coupled into the resonator in the backward direction. Photodiodes monitor the input, transmission and return powers of the laser, while the laser power is adjusted with a variable attenuator. Circ = fibre-optical circulator, PD = photodiode.}
    \label{fig:fig1}
\end{figure}

%\begin{refsection}
Nonreciprocal light propagation \cite{Potton2004} is expected to play an important role in future photonic networks and optical data processing. A challenge for integrated photonic circuits is the realization of nonreciprocal elements like isolators and circulators \cite{Jalas2013}. Most optical isolators and circulators are based on the magneto-optic Faraday effect, which requires the integration of (electro)magnets \cite{bi2011chip,Dotsch2005,Kono2007,Shoji2008,Pintus2017}. Additional and compelling ways to achieve optical nonreciprocity include specially designed waveguides \cite{Feng2011,Gallo2001}, optomechanical systems \cite{Fang2017,Manipatruni2009,Ruesink2016,Shen2016,Wang2015}, parity-time symmetry breaking in resonator systems \cite{Chang2014,Peng2014}, indirect interband photonic transitions \cite{Lira2012,Yu2009}, Brillouin scattering \cite{Huang2011,Kang2011}. Other systems include spatially asymmetric “optical diodes” based on microresonator systems \cite{Fan2011}. This device works based on asymmetric thermal resonance frequency shifts, with the disadvantage that it would suffer from dynamically reciprocal backwards transmission in the spectral vicinity of the forward transmitted light \cite{Shi2015}. 
In this work we demonstrate a novel and simple isolator and circulator based on the intrinsic nonreciprocity of the Kerr nonlinearity in a single whispering gallery microresonator. The underlying Kerr-nonreciprocity corresponds to the physical effect that counterpropagating light induces twice the Kerr shift compared to the same amount of co-propagating light \cite{Agrawal2006,Kaplan1982}. In a passive microresonator this leads to spontaneous symmetry breaking between clockwise and counterclockwise modes, which has been recently demonstrated \cite{DelBino2017,Cao2017}. Here, we make use of this Kerr-nonlinearity-induced nonreciprocity in a microresonator in an add-drop configuration to realize a novel type of optical isolator and circulator \cite{Pintus2013}. In contrast to the case of spontaneous symmetry breaking, we intentionally set a microresonator into a state where it can only support unidirectional propagation of light at a given frequency. Our measurements of the power scaling of the isolation are in excellent agreement with theoretical calculations based on the interplay of the Kerr nonlinearity with the modes of an ultra-high-Q microresonator. 
Sending equal amounts of light in counterpropagating directions into a passive microresonator with $\chi^{(3)}$ nonlinearity  leads to spontaneous symmetry breaking of the clockwise (cw) and counterclockwise (ccw) modes \cite{DelBino2017,Cao2017}. In addition, this effect induces optical nonreciprocity such that the cw and ccw modes of a microresonator acquire different resonance frequencies, as shown in \fref{fig:fig1}(a).

In the biased state (with input laser on) the coupled resonator breaks Lorentz reciprocity and as a consequence, light at the input frequency can only circulate in one direction. Only backward propagating light (or noise) that falls within the spectrally well-separated clockwise resonance could get back to the input port (obeying dynamic reciprocity \cite{Shi2015}). If desired, this narrow backwards transmitting window can be closed by using a second isolator with slightly different mode splitting in series. Once the resonator is in a symmetry-broken state with unidirectional light propagation, the power sent in the backwards direction towards the resonator can even exceed the forward propagating power without changing the state of the resonator \cite{DelBino2017}.
\\

In our experiments we use a fused silica microrod resonator ($Q$-factor ${=1.5\times10^8}$, diameter = 1 mm) coupled to two tapered optical fibers in an add-drop configuration (see setup in \fref{fig:fig1}(b)). Light from a diode laser at 1550 nm is sent into the microresonator in the ccw direction. The microresonator mode passively locks itself to the laser frequency, making the system insensitive to laser frequency drifts \cite{Carmon2004}. The light is then coupled out of the resonator through the second tapered fiber and directed to an optical circuit of interest. Any reflection from this circuit cannot couple back into the resonator because it is not resonant in the clockwise direction, and is thus transmitted through the tapered fiber to an absorber. A common application for optical isolators is to prevent back-reflections in optical circuits from disturbing the operation of a laser. One main advantage of optically induced isolation via the Kerr effect is the increase in isolation with optical power (as higher power leads to a larger frequency splitting between the cw and ccw resonator modes). We analyze the power scaling of the isolation and simulate a worst-case scenario where all of the transmitted power is reflected back to the resonator by using a fiber mirror. The behaviour of the setup is characterized with photodiodes that monitor the input, transmitted, and reflected optical power levels (see \fref{fig:fig1}(b)).

\begin{figure}[h]
    \centering
    \includegraphics[width=\textwidth]{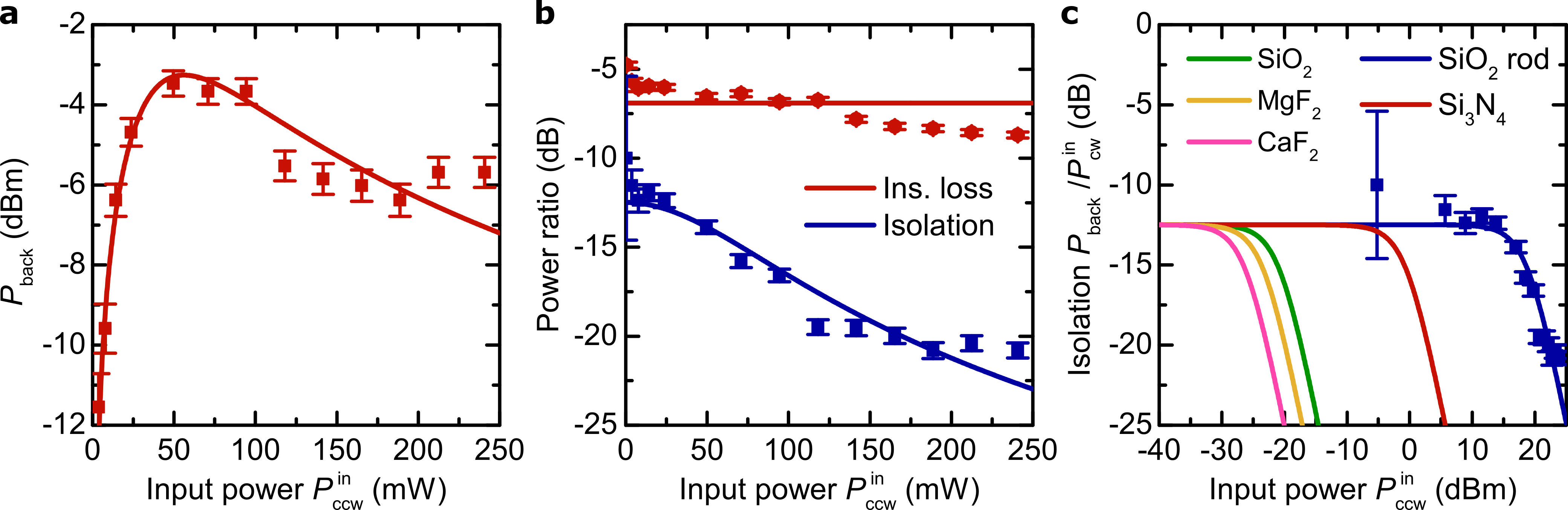}
    \caption{Kerr-nonlinearity-induced isolation at 1550 nm, theory and experiment. (a) Measurement of return power $P\mathrm{_{back}}$  vs. input power $P\mathrm{_{ccw}^{in}}$  with theoretical fit. For low input powers, the return power is proportional to the input power as expected for a linear system. For high powers, the return power decreases due to the nonreciprocal Kerr effect. (b) Insertion loss $P\mathrm{_{out}}/P\mathrm{_{ccw}^{in}}$ and isolation $P\mathrm{_{back}}/P\mathrm{_{cw}^{in}}$  vs. input power. We measure an insertion loss of around 7 dB and a maximum isolation in excess of 24 dB. (c) Predicted isolation for waveguide resonators of various materials, assuming an effective mode cross-sectional area of 4 $\mathrm{\mu m}^2$ and a resonator diameter of 100 $\mathrm{\mu m}$. Our measurement ($\mathrm{SiO_2}$ rod resonator, diameter 1 mm) is included for comparison. The respective Q-factors for the calculations are $10^9$ for magnesium fluoride and calcium fluoride \cite{Grudinin2009,Savchenkov2015}, $5\times 10^7$ for silicon nitride \cite{Xuan2016} and $5\times 10^8$ for $\mathrm{SiO_2}$.  Our calculations show isolation at sub-milliwatt power levels and even down to tens of microwatts in these currently available microresonator systems.}
    \label{fig:fig2}
\end{figure}

\Fref{fig:fig2} shows the experimental results for insertion loss and isolation versus the launched optical power. The power $P\mathrm{_{ccw}^{in}}$ corresponds to the power entering the input tapered fiber, $P\mathrm{_{out}}$  is the power exiting the second tapered fiber, and $P\mathrm{_{back}}$  is the power that passes through the resonator in the backwards direction. The data in \fref{fig:fig2}(a) shows an initial increase in backward-propagating light with input power. At the point where the Kerr-nonreciprocity starts to split the cw and ccw resonance frequencies, the isolation kicks in and reduces the backward-propagating power, which can be seen at around 10 mW of input power in \fref{fig:fig2}(a). At higher powers, the optical isolation reaches -24 dB (\fref{fig:fig2}(b)). The insertion loss of the setup (transmission in the forward direction) is around -5 dB. Note that the threshold powers is comparably high as a result of the large diameter and large mode volume of the resonators in these proof-of-principle experiments. The threshold powers could be significantly reduced and the attainable isolation improved by using smaller resonators. This is illustrated in \fref{fig:fig2}(c), which shows the calculated isolation in resonators made from different materials.

The theoretical curves in \fref{fig:fig2} are calculated using a model that takes into account the interaction between counterpropagating light via the Kerr nonlinearity. The optical powers $P\mathrm{_{cw}}$ and $P\mathrm{_{ccw}}$ coupled into the resonator in the clockwise and counterclockwise directions are given by the following coupled equations \cite{DelBino2017}:
\begin{equation}
P\mathrm{_{ccw}}=\frac{\eta\mathrm{^{in}}P\mathrm{_{ccw}^{in}}}{1+\left(\dfrac{\delta}{\gamma}+\dfrac{1}{P_0}\left(P\mathrm{_{cw}}+2P\mathrm{_{ccw}}\right)\right)}
\end{equation}
\begin{equation}
P\mathrm{_{cw}}=\frac{\eta\mathrm{^{in}}P\mathrm{_{cw}^{in}}}{1+\left(\dfrac{\delta}{\gamma}+\dfrac{1}{P_0}\left(P\mathrm{_{ccw}}+2P\mathrm{_{cw}}\right)\right)}
\end{equation}

Here, $P\mathrm{_{ccw}^{in}}$ and $P\mathrm{_{ccw}^{in}}$ are the powers launched in counterpropagating directions into the coupling waveguides, $\eta\mathrm{^{in}}$ is the in-coupling efficiency from the waveguides to the resonator, and $\delta$ is the detuning of the laser frequency from the cold cavity resonance with a half-linewidth $\gamma$. We use a symmetrically coupled resonator, which means that $\eta\mathrm{^{in}}=4k(\gamma-k)/\gamma^2$ is the same for both waveguides, with $k$ being the coupling coefficient between the resonator and each waveguide. The terms $\left(P\mathrm{_{ccw}}+2P\mathrm{_{cw}}\right)/P_0$ and $\left(P\mathrm{_{cw}}+2P\mathrm{_{ccw}}\right)/P_0$ correspond to the normalized nonreciprocal Kerr resonance frequency shifts. Important for the isolator/circulator to work is the factor of 2 in the dependence on the respective counterpropagating power, which enables the isolation via interaction of counterpropagating light in a single resonator (compared to previous work on coupled microresonator systems\cite{Fan2011}). $P_0$ is a resonator-specific constant given by 

\begin{equation}
P_0=\frac{\pi^2n_0^2dA\mathrm{_{eff}}(\gamma-k)}{Q\mathrm{_{L}}^2n_2\lambda\gamma}
\label{eq:p0}
\end{equation}

for refractive index $n_0$ at wavelength $\lambda$, nonlinear refractive index $n_2$, resonator diameter $d$, effective mode area $A\mathrm{_{eff}}$ and loaded quality factor $Q\mathrm{_{L}}$. The output and return powers are given by $P\mathrm{_{out}}=\eta\mathrm{^{out}}P\mathrm{_{ccw}}$ and $P\mathrm{_{back}}=\eta\mathrm{^{out}}P\mathrm{_{cw}}$  with the out-coupling coefficient $\eta\mathrm{^{out}}=k/(\gamma-k)$. Assuming that the laser is on resonance with the ccw mode, the insertion loss $P\mathrm{_{out}}/ P\mathrm{_{ccw}^{in}}$ will be equal to the through-coupling efficiency $\eta\mathrm{^{thru}}=\eta\mathrm{^{in}}\eta\mathrm{^{out}}=4k^2/\gamma^2$. Assuming that the rest of the optical circuit has a reflectivity $R$ such that $P\mathrm{_{cw}^{in}}=RP\mathrm{_{out}}$, we thus find that the isolation $I=P\mathrm{_{back}}/P\mathrm{_{cw}^{in}}$ is given by

\begin{equation}
I=\frac{\eta^\mathrm{thru}}{1+\left(\dfrac{P\mathrm{_{cw}^{in}}\eta\mathrm{^{in}}}{P_0}(1-IR)\right)^2}
\end{equation}

Note that the insertion loss of the microresonator-isolator system can be decreased by increasing the coupling to the tapered fibers. However, this would lead to a decreased Q-factor and a trade-off between the attainable isolation and insertion loss.

The resonance frequency difference in our system is of the order of 10 linewidths corresponding to 50 MHz and could be increased to GHz ranges by using microresonators with higher optical nonlinearity or by operating higher above the threshold power. The resonance frequency difference in terms of cavity linewidths is comparable to existing electromagnetic microresonator systems with larger absolute resonance frequency differences \cite{Huang2016}. In terms of applications our system could be used to isolate chip-integrated continuous wave laser diodes (with subsequent modulation if used in telecom systems).

\begin{figure}[h]
    \centering
    \includegraphics[width=\textwidth]{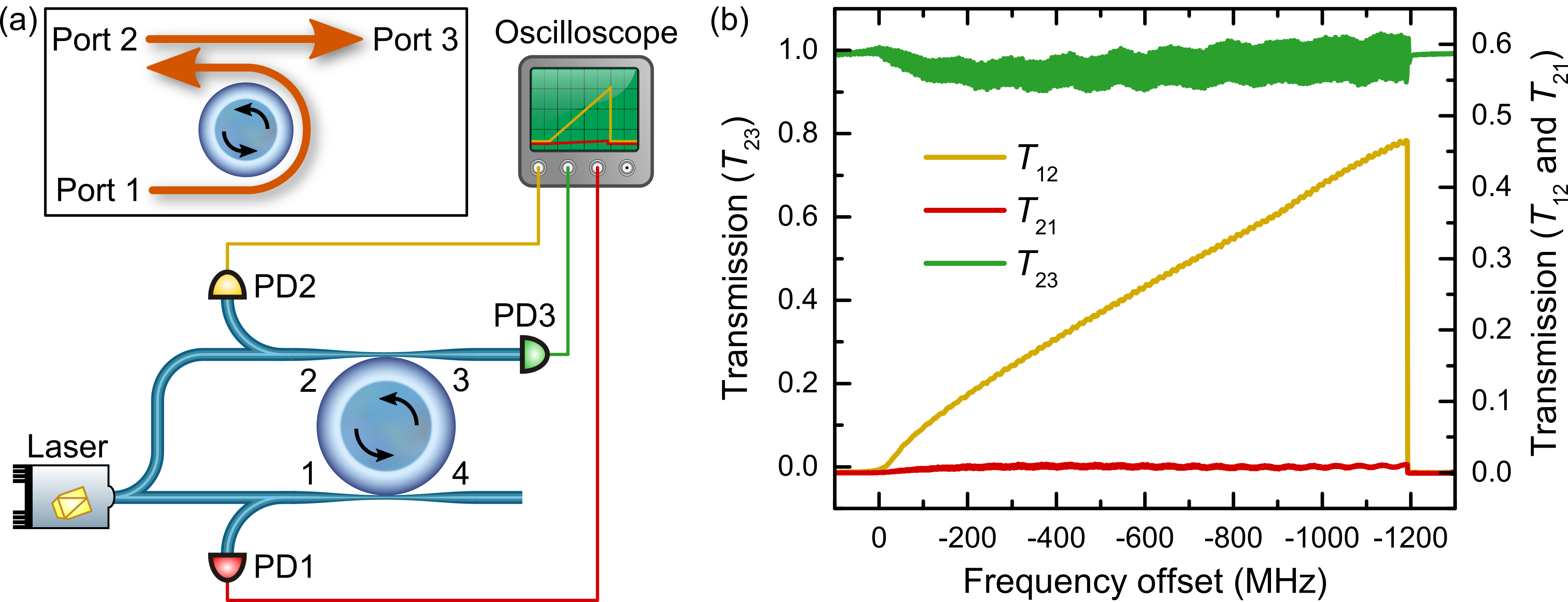}
    \caption{Microresonator-based circulator. (a) Schematic of the setup. Light is split between ports 1 and 2 with slightly more light going into port 1, so that the resonator is set into the counterclockwise (ccw) state. Photodiodes measure the optical output powers from each port. (b) Transmission $T_{ij}$ (from port i to port j) as the laser frequency is swept through the resonance.}
    \label{fig:fig3}
\end{figure}

In addition to operating as an isolator, a microresonator in an add-drop configuration can be used to realize a three-port circulator \cite{Pintus2013} as shown in \fref{fig:fig3}(a). Light sent into port 1 is transmitted via the resonator to port 2. The light circulating counterclockwise around the resonator causes the clockwise resonance frequency to be shifted away. This means that light of the same frequency entering port 2 cannot couple into the resonator and is instead transmitted to port 3. This was tested by splitting light from a laser between port 1 (404 mW) and port 2 (87 mW), and measuring the output powers with fast photodiodes. The transmission coefficients, i.e. the output powers normalized by the relevant input power, are plotted in \fref{fig:fig3}(b) while sweeping the laser frequency across the resonance.
\\

In our microresonator-based circulator we observe 47\% transmission from port 1 to port 2 (corresponding to 3.3 dB insertion loss) and a directivity $T_{23}/T_{21}$ of $>$19 dB while the microresonator is thermally locked to the laser. The oscillations in $T_{23}$ result from interference with light from port 1 that is back-scattered within the resonator. From the amplitude of these oscillations, we obtain a value of -33 dB for the (unwanted) transmission coefficient $T_{13}$. This is comparable with values for commercial magneto-optic circulators, and may be further improved by reducing the backscattering through surface defects of the resonator, which will also increase the Q-factor and thereby the isolation. In order to achieve stable circulation for larger laser frequency changes than those supported by the thermal self-locking of the microresonator, the resonator could be temperature controlled and/or the laser could be injection locked to the resonance.

In conclusion, we have demonstrated an optical isolator and circulator based on the intrinsic nonreciprocity of the Kerr effect in an optical microresonator. The biased microresonator breaks Lorentz reciprocity and our results show increasing isolation with optical power which is supported by the theoretical model based on the nonreciprocal $\chi^{(3)}$-nonlinearity. In our experiments we achieve more than 20 dB isolation. The threshold power in our millimetre-sized rod resonators is around 10 mW, which could be significantly reduced to tens of microwatts by using smaller resonators with higher finesse and higher nonlinearity. In chip-based optical devices, such isolators would be ideal for reducing back-reflections into integrated laser diodes, especially given the increasing isolation at higher optical powers. Aside from the fundamental physics of the intrinsic Kerr nonreciprocity in ring resonators, our work provides a new and simple route for realizing optical isolators and circulators for applications in integrated photonic circuits.

\subsection*{Acknowledgements}
This work was supported by the National Physical Laboratory’s Strategic Research Programme and the Marie Sklodowska-Curie Grant Agreement No. 748519 CoLiDR. This project has received funding from the European Research Council (ERC) under the European Union’s Horizon 2020 research and innovation programme (grant agreement No 756966). XZ acknowledges support from the Chinese Scholarship Council and Natural Science Foundation of China grants 61435002, 61675015. JMS acknowledges funding via a Royal Society of Engineering fellowship. LDB and MTMW acknowledge funding from EPSRC through the Centre for Doctoral Training in Applied Photonics.
\\

$\dagger$ these authors contributed equally to this work.
\\

See Supplement 1 for supporting content. 
\pagebreak
%\printbibliography[heading=subbibliography]
%\end{refsection}

\pagebreak

\bibliographystyle{ieeetr}
\bibliography{Leo2}
\pagebreak

\appendix
%\begin{refsection}
\section*{Supplementary information}
\subsubsection*{Simulation of other materials}
In the simulations for other materials we used a fixed diameter of 100 $\mathrm{\mu}$m, and assumed the same taper losses and mirror reflectivity as in our experiment. The values chosen here are taken from other recent works and are only representative of the expected performance level. In general, a smaller mode volume and higher Q-factor are required to improve the efficiency. At the same time, reflections inside the resonator must be avoided.

\begin{table}[h]
\centering
\label{tab:parameters}
\begin{tabular}{|l|l|l|l|l|}
\hline
Material             & $A_\mathrm{eff}$     & $n$\cite{Polyanskiy}   & $n_2$                               & $Q_0$                          \\ \hline
$\mathrm{SiO_2}$ rod & $10\,\mathrm{\mu m^2}$ & 1.444 & $2.7\times 10^{-16}\,\mathrm{cm^2/W}$ \cite{Kato1995}& $1.5\times10^8$ \\ \hline
$\mathrm{Si_3N_4}$ \cite{Xuan2016}  & $4\,\mathrm{\mu m^2}$  & 2.463 & $2.4\times 10^{-15}\,\mathrm{cm^2/W}$ & $5\times10^7$   \\ \hline
$\mathrm{SiO_2}$     & $4\,\mathrm{\mu m^2}$  & 1.444 & $2.7\times 10^{-16}\,\mathrm{cm^2/W}$ \cite{Kato1995}& $5\times10^8$   \\ \hline
$\mathrm{CaF_2}$ \cite{Savchenkov2013}    & $4\,\mathrm{\mu m^2}$  & 1.426 & $1.9\times 10^{-16}\,\mathrm{cm^2/W}$ & $1\times10^9$   \\ \hline
$\mathrm{MgF_2}$ \cite{Herr2013}   & $4\,\mathrm{\mu m^2}$  & 1.37  & $9\times 10^{-17}\,\mathrm{cm^2/W}$   & $1\times10^9$   \\ \hline
\end{tabular}
\caption{Parameters used to generate figure 2(c) on the main article}
\end{table}

Additional literature reports of $Q$ up to $10^{10}$ for $\mathrm{CaF}_2$ \cite{Hofer2010} and similar values have been attained in $\mathrm{MgF}_2$. These resonators, however, have larger diameters, which would increase the nonlinear threshold power.

\subsubsection*{Frequency splitting}
The isolation of our device relies on the fact that it exhibits two different resonance frequencies for light propagating in the two directions. The value of the frequency shift depends on the power launched in the isolator and reflected by the test circuit. We cannot measure this shift directly in our current setup. However, it is possible to calculate it from the coupled modes equations (Eq. (1) and (2) in the main article) using the value of $P_0$ fitted by the measured data and the circulating powers measured by the photodiodes. In the case of the circulator the frequency splitting $\Delta \omega$ between the resonances in the two directions satisfies:
\begin{equation}
\frac{\Delta \omega}{\gamma} = \frac{P_\mathrm{ccw}-P_\mathrm{cw}}{P_0}=\frac{190 \mathrm{mW}-0\mathrm{mW}}{10.8\mathrm{mW}}=18
\end{equation}
where $P_\mathrm{ccw}$ and $P_\mathrm{cw}$ are the coupled powers measured as the drops in the transmission of the tapered fibres with respect to being out of resonance. The loaded $Q$-factor $Q_\mathrm{L}$ with two tapered fibres coupled to the resonator is about $3\times 10^7$ (to be compared with the raw $Q$-factor of $1.5 \times 10^8$) corresponding to a half linewidth $\gamma$ of about $2\pi \times 3$ MHz. Therefore we can calculate a frequency splitting of 54 MHz for this particular combination of microresonator and input powers. However this value strongly depends on the characteristic power for the isolation to occur (see equation (3) in the main article).
%:
%\begin{equation*}
%P_0=\frac{\pi^2n_0^2dA\mathrm{_{eff}}(\gamma-k)}{Q\mathrm{_{L}}^2n_2\lambda\gamma}
%\end{equation*}
Thus, it strongly depends on the $Q$-factor of the resonator and nonlinear refractive index $n_2$ of the material.
%\printbibliography[heading=subbibliography]
%\end{refsection}

\end{document}